\documentclass[prd,superscriptaddress,onecolumn]{revtex4-1}
\usepackage{color}
\usepackage{bm}
\usepackage{graphicx}  
\usepackage{dcolumn}   
\usepackage{bm}        
\usepackage[english]{babel}
\usepackage{booktabs,multirow,array}
\usepackage[bottom]{footmisc}
\usepackage{dblfnote}
\usepackage{mathtools}
\usepackage{amsmath}
\usepackage{physics}
\usepackage{braket}
\DFNalwaysdouble 
\usepackage[colorlinks]{hyperref}
\definecolor{LinkColor}{rgb}{0.75, 0, 0}
\definecolor{CiteColor}{rgb}{0, 0.5, 0.5}
\definecolor{UrlColor}{rgb}{0, 0, 0.75}
\hypersetup{linkcolor=LinkColor}
\hypersetup{citecolor=CiteColor}
\hypersetup{urlcolor=UrlColor}
\begin{document}
\title{Selective enhancement of quantum decay channels}
\author{Pritam Nanda}
\email{pritam@iisermohali.ac.in}
\affiliation{Department of Physical Sciences,
IISER Mohali, Sector 81 SAS Nagar, Punjab, India.}
\author{Kinjalk Lochan}
\email{kinjalk@iisermohali.ac.in}
\affiliation{Department of Physical Sciences,
IISER Mohali, Sector 81 SAS Nagar, Punjab, India.}


\begin{abstract}
In the decay of quantum particles under field theoretic consideration, the decay rate is typically  a convolution of the  density of modes the  primary field is  allowed to decay into and the allowed probability density for the field to decay into such modes. In free space, though many such processes show high amplitude of such transitions towards the infrared sector, the depletion of allowed mode density in that regime arrests the efficacy of such decays at low energies. Therefore in free space, in order to enhance the decay rate, one needs the transition probability density to be rich enough towards the high energy sector where mode density support is also high enough to make the rate sufficiently large. In this work we argue that in the controlled boundary condition environment e.g. in a cavity, the mode functions of product field receive significant support towards their infrared sector,  boosting the probability (and hence rates) of low energy processes. The cavity geometry offers sweet spots in terms of resonant geometry around which the interaction of a primary field with product fields receives dramatic enhancement, significantly enlarging its decay rates. Therefore, a judicious selection of cavity geometry serves as a potential substitute to studying interesting processes at high energy. The results have direct relevance for the study of QED processes and implications for the study of exotic new physics are also discussed.
 \end{abstract}
\maketitle

\section{Introduction}
 Quantum field theory (QFT) comes up with many intriguing phenomenon involving fundamental fields at the quantum level. Interacting fields display many intriguing features like particle decay \cite{Andreassen}. Decay rate of particles is one of the fundamental predictions of the QFT and an observation of the decay process is one of the powerful probe of learning about interaction and properties of quantum fields at different energy scales  \cite{Srednicki_2007}. 
 
 Many subtle issues regarding QFT are contained in the scattering matrix elements which characterize the transition between an in-state to possible out-going states. Different fields interacting with each other modify the spectrum of outgoing states compared to the incoming states, through the interaction.  These shifts in the quantum state depend upon strength of interaction and permissibility of processes within the domain of interacting quantum fields, and are also dictated by the symmetries of the theory \cite{Audretsch_1985, Audretsch_1986, Audretsch_1987}. The Ward-Takahashi identities directly constrain the possibility space of outcomes consistent with the symmetries in the system  \cite{Ward:1950,Takahashi:1957xn,Nakanishi:1974pz}. 

In the phenomenological setting, the imprints of any particular process is estimated through its potential signatures in the out-state.  This becomes of primary importance in  the estimation of decay channels of an unstable particle. However this is an ambitious exercise as there is always a hierarchy of strength of decay channels \cite{Giacosa:2011xa}. Usually the most strongly coupled channel dominates the out state in any scattering or decay process \cite{Khoze:2017tjt}.  Therefore, to isolate and decipher, let us say, the marginal contributions of an intriguing but subdominant decay channel is  typically a herculean task.   The usual way of uncovering such characteristics is to keep studying this process at higher and higher energies where the contributions of such secondary channels become somewhat larger and observable against the background (of other anticipated decay channels), or into the regime where the higher order contributions of such decay channels are qualitatively distinguishable \cite{Crivellin_2011,Dubovyk:2018rlg}. Even in this program, ultimately what we really study is the deviation of the observed data from the prediction of the primary decay channel. Thus, in such studies we end up constraining the collective contribution of all subdominant processes. As we shall see below, for a particular decay channel, the rate of decay in the energy domain of the product field modes can be expressed as 
\begin{equation}
    \Gamma_{|in\rangle \rightarrow \sum |final\rangle}   \sim  \lambda^2 \int d\omega_k \rho(\omega_k) {\cal K}(\omega_k, \{Q\}) +  {\cal O} (\lambda^4)\,,
\end{equation}
where $\lambda$ is the coupling of interaction, $\rho(\omega_k)$ is density of modes of the quantum field, the primary field decays into, ${\cal K}(\omega_k, Q)$ is the contribution function (obtained via the transition amplitude) of modes in the decay process, while $\{Q\}$ signifies conserved quantities in the process.  While the function $\rho(\omega)$ is sensitive to the mode density of the  product field, the function  ${\cal K}(\omega_k, \{Q\})$ depends upon the process details e.g. the interaction term, the states of interest etc.

In any typical field interaction process, both these functions determine the effectivity of a decay channel. In free space,  the number of modes is small for infrared sector ($\rho(\omega_k) \sim {\cal  O}(\omega_k^2)$)  and thus the effective contribution becomes diluted and process is dominated by high energy modes. 
In some recent works \cite{Stargen:2021vtg}, though, it is argued that in the interaction of fields with matter, the density of states for the field modes in the energy space, participating in the process, can be engineered to play a crucial role. Inside a cavity this density of modes can be modified suitably to provide a strong support at the resonance frequency.  Thus if the infrared modes play significant role in a scattering process, modifications of boundary conditions can potentially enhance them suitably  by populating the density of modes towards the infrared sector suitably.  In this work, we show that using the judicious boundary conditions on the participating fields we can potentially and selectively enhance some of the decay channels in quantum field theory, despite them not being the primary channel in terms of coupling strength in free space. Using appropriate mode functions consistent with boundary conditions, one can turn even the weakest channels into the most prominent route of decays with very specific boundary conditions. We demonstrate this concept through the study of processes involving scalar fields, but the ideas developed here are quite general in nature and can be applied to any field in which we have control over the boundary conditions, in a particular process.

\section{Decay of a scalar particle in free space (LEADING ORDER
CALCULATION)}\label{Section_II}
For demonstration, we consider the decay of a scalar field state into the particles of other scalar fields (though this can be generalized to have more realistic interaction with other kind of fields as well).
Let us examine the following Lagrangian, which involves real scalar field, $\Phi$ which can decay either into another field $\phi_1$ or a field $\phi_2$ \cite{Giacosa:2011xa} 

\begin{eqnarray} \label{IntLang}
{\cal L}  = \frac{1}{2} \left[-\partial_{\mu} \Phi  \partial^{\mu} \Phi - M^2 \Phi^2  -\partial_{\mu} \phi_1  \partial^{\mu} \phi_1- m_1^2 \phi_1^2 -\partial_{\mu} \phi_2  \partial^{\mu} \phi_2 - m_2^2 \phi_2^2 \right] -\lambda_1  \Phi  \phi_1  \phi_1  -\lambda_2  \Phi  \phi_2  \phi_2 \,.
\end{eqnarray}
Since the decay of the field $\Phi$ into either of the fields is exclusive, the description of decay into one of the fields can be studied via an effective Lagrangian
\begin{equation}
    \mathcal{L_{\text{eff}}}=-\frac{1}{2}\partial^\mu\Phi\partial_\mu\Phi-\frac{1}{2}M^2\Phi^2  - \frac{1}{2}\partial^\mu\phi\partial_\mu\phi-\frac{1}{2}m^2\phi^2-\lambda\Phi\phi\phi \,.
\end{equation}
The final term in the Lagrangian represents an interaction between the two fields.  With the field definitions 
\begin{eqnarray}
\Phi \text{ or } \phi(x) &=& \int d^3\vec{k}\, \left(\hat{a}_{\vec{k}}\,  u_{\vec{k}}(x)  + \hat{a}_{\vec{k}}^{\dagger}  u_{\vec{k}}^*(x)\right), \qquad \text{where,}\\
u_{\vec{k}}(x)
&=&
\frac{e^{ik\cdot x}}
{\sqrt{(2\pi)^3\,2\omega_{\vec{k}}}} . 
\end{eqnarray}
where
\begin{equation}
[\hat{a}_{\vec{k}},\hat{a}_{\vec{k}'}^{\dagger}] = \delta^3(\vec{k}-\vec{k}')
\end{equation}
and $\omega_{k}^2= {k_z}^2+k_x^2 + k_y^2 + M^2 \text{~~or~~} m^2,$ for the two fields, we can proceed to calculate the lifetime of the $\Phi$ particle under a decay process. We consider a process in which a $\Phi$ particle $|\vec{P}_{\Phi}\rangle =  \sqrt{(2\pi)^3~2\omega_{\vec{P}_{\Phi}}}\hat{a}_{\vec{P}_{\Phi}}^{\dagger}\ket{0}_{\Phi}$ decays into two $\phi$ particles $|\vec{p}_1,\vec{p}_2\rangle  =  1/2 \times \sqrt{(2\pi)^3~   2\omega_{\vec{p}_1}}\sqrt{(2\pi)^3~2\omega_{\vec{p}_2}}\hat{a}_{\vec{p}_1}^{\dagger} \hat{a}_{\vec{p}_2}^{\dagger}\ket{0}_{\phi_1}\ket{0}_{\phi_2}$.
The decay probability at the tree level, is defined as,
\begin{equation}   \label{ProbT}
  {\cal P}_{|in\rangle \rightarrow \sum |final\rangle}   \approx \int \frac{d^3\vec{p_1}}{(2\pi)^32E_{p_1}}\int \frac{d^3\vec{p_2}}{(2\pi)^32E_{p_2}} \frac{~|-i\lambda\bra{\vec{p_1} \vec{p_2}}\int d^4x\hat{\Phi}(x)\hat{\phi}_1(x)\hat{\phi}_1(x)\ket{\vec{P_{\Phi}}}|^2}{\langle \vec{P}_{\Phi}|\vec{P}_{\Phi} \rangle }
 \end{equation}
Using the plane wave modes $e^{i k \cdot x}$ in free space for both the fields, the amplitude  for this process to the lowest order in $\lambda$,  can be written as

\begin{eqnarray}
   \bra{\vec{p_1} \vec{p_2}}\int d^4x\hat{\Phi}(x)\hat{\phi}_1(x)\hat{\phi}_1(x)\ket{\vec{P}_{\Phi}} =\int d^4x e^{i(P_\Phi-p_1-p_2)\cdot x}  = (2\pi)^4 \delta^4(P_{\Phi}-p_1-p_2) \nonumber\\
   =  (2\pi)^4 \delta(E_{\vec{P}_\Phi}-E_{\vec{p_1}}-E_{\vec{p_2}})\delta^3(\vec{P}_{\Phi}-\vec{p_1}-\vec{p_2}),\\
\end{eqnarray}
which is  effectively the statement of the symmetries in the Lagrangian, i.e. the conservation of the 4-momentum in the interaction at the tree level. 
Now if we do the computation in the co-moving frame of the in-particle\footnote{This trick however, can not be employed for the massless particle decay}, where $\vec{P_\Phi}=0,~ E_{\vec{P_\Phi}} = M$, we can write the {\it decay rate density}  $\rho_{\Phi \rightarrow \phi \phi}$ (see Appendix \ref{appendix_A})  as 

\begin{equation}
    {\cal P}_{|in\rangle \rightarrow\sum |final\rangle}  = \int dt \int d^3 x \underbrace{\left[\frac{2\pi}{\delta^{3}(0)} \frac{\lambda^2}{2E_{\vec{P}_\Phi}}  \int\frac{d^3\vec{p}_1d^3\vec{p}_2}{2E_{\vec{p}_1}2E_{\vec{p}_2}} e^{i(P_\Phi-p_1-p_2)\cdot x}~\delta^4(P_\Phi-p_1-p_2)\right]}_{\rho_{\Phi \rightarrow \phi \phi}(x)}\,.  \label{FiniteRegionRate}
\end{equation}
Using this density of decay rate, we can obtain the rate of decay in any finite region of space as well
\begin{equation}
   \Gamma_V = \left[\int_V d^3 x \rho_{\Phi \rightarrow \phi \phi}\right] =  \left[\frac{2\pi V}{\delta^{3}(0)} \frac{\lambda^2}{2E_{\vec{P}_\Phi}}\int\frac{d^3\vec{p}_1d^3\vec{p}_2}{2E_{\vec{p}_1}2E_{\vec{p}_2}}\delta^4(P_\Phi-p_1-p_2)\right]\,.  \label{FiniteRegionRate}
\end{equation}
The factor $V/\delta^{3}(0)$ in the expression of the decay rate $\Gamma_V$, marks the factor for decay in a subregion  of volume $V$ in the case of a spatially uniform probability of decay, obtainable for a momentum eigenstate $\ket{P_{\Phi}}$ as the in-state. The rest frame (of the decaying particle) expression of  the rate of decay in all space $(V/(2\pi)^3\delta^3(0) =1)$ is
\begin{equation}\label{equation_5} 
    \Gamma_0  \stackrel{V \rightarrow \infty}{=} \frac{\lambda^2}{M}\int\frac{d^3\vec{p_1} d^3\vec{p_2}}{(2\pi)^2}\frac{1}{4E_{\vec{p_1}}E_{\vec{p_2}}}\delta(M-E_{\vec{p_1}}-E_{\vec{p_2}})\delta^3(\vec{p_1}+\vec{p_2}).
\end{equation}
Integration with respect to $\vec{p}_2$ gives
\begin{equation}
    \Gamma_0=\frac{\lambda^2}{M}\int\frac{d^3\vec{p_1}}{(2\pi)^2}\frac{1}{4E_{\vec{p_1}}^2}\delta(M-2E_{p_1}) = \frac{\lambda^2}{M}\int\frac{d^3\vec{p_1}}{(2\pi)^2}\frac{1}{4(\vec{p_1}^2+m^2)}\delta\big(M-2\sqrt{\vec{p_1}^2+m^2}\big),
\end{equation}
using the dispersion relation for $\phi$. 
We now employ the identity for the Dirac delta function: $\delta(f(p)) = \sum_i \frac{\delta(p - p_i)}{|f'(p_i)|}$. Additionally, in momentum space, the volume element is expressed as $d^3\vec{p} = 4\pi p^2dp$.  Going to the energy domain of the product field $\phi$  we obtain
\begin{eqnarray}\label{eq_8}
     \Gamma_0=\frac{\lambda^2}{ M}\int\frac{4\pi p^2 dp}{(2\pi)^2}\frac{1}{4(\vec{p_1}^2+m^2)}\delta\big(M-2\sqrt{\vec{p_1}^2+m^2}\big)
      =\lambda^2\int_0^\infty ~ d\omega \underbrace{\Theta(\omega-m)\frac{4\pi \omega \sqrt{\omega^2 -m^2}}{(2\pi)^2 M}}_{\rho(\omega)}\underbrace{\frac{1}{4\omega^2}\delta\big(M-2\omega\big)}_{{\cal K}(\omega, M)}.
\end{eqnarray}
Here we have defined the rate  as the convolution between the density of modes  $\rho(\omega)$ of $\phi$ and the transition probability ${\cal K}(\omega, M)$ for the decay into a mode with energy $\omega$. Through ${\cal K}(\omega, M)$ one can see that in the energy domain only those modes participate in this process which have half the energy of the decaying particle.  One can further see that due to $\omega^2$ in the denominator, the modes of low energies (hence decay  of low mass particles) would impart a large contribution  into the process. Thus {\it prima facia} one could have anticipated that decay of low mass particles  should have been prevalent in nature. Yet, it is the dilution of $\rho(\omega) \sim \Theta(\omega-m) \omega \sqrt{\omega^2 -m^2} \sim {\cal  O}(\omega^2)$ at low frequencies which ultimately regulates the decay rate at low frequencies for low mass particles.
A straightforward calculation then yields the decay rate as \footnote{For the ease of computation, we had selected the in-state as $|P_{\Phi} \rangle$ with three momentum $\vec{P}$, and then we obtained the rate in the rest frame of this in-particle. In general a more spread out state can be chosen 
$$|in \rangle = \int \frac{d^3\vec{P}}{2E_{\vec{P}}} \tilde{\Phi}(\vec{P})|\vec{P} \rangle,$$
which does not change the results obtained qualitatively  (see Appendix \ref{appendix_A}).} :
\begin{equation}
    \Gamma_0=\Theta(M-2m) \frac{\lambda^2}{8\pi M}\sqrt{1-\frac{4m^2}{M^2}}. \label{GDR}
\end{equation}
Further, the decay rates into the two fields of Eq.~(\ref{IntLang}) can be compared to be 
\begin{eqnarray}
\frac{\Gamma_{\Phi  \rightarrow \phi_1}}{\Gamma_{\Phi  \rightarrow \phi_2}} = {{\lambda_1^2 ~\Theta(M-2m_1)\sqrt{1-\frac{4m_1^2}{M^2}} \over \lambda_2^2 ~\Theta(M-2m_2)\sqrt{1-\frac{4m_2^2}{M^2}}}}.
\end{eqnarray}
Therefore, in standard situation the prominent decay channel will primarily be the one whose coupling is stronger. Their relative strengths will decide their contribution in the decay channels. To observe the relevance  of  a subdominant channel, one would  need to employ  techniques increase  its  visibility by moving  to high energies, if the couplings run differently with energies.  
In what follows we propose a boundary condition controlled decay rate for one of the channels.  One of the decaying (let  us say $\phi_1$ ) modes are contained in a cavity environment via putting in Dirichilet boundary conditions on it. This modifies the symmetries of the system and therefore the decay rate computation can be expected to be corrected. We argue that suitable geometry selection can compensate for requirement of energy enhancement for enhancing a particular channel's contribution.  

\section{Tree level Decay of scalar particle inside cavity  region}

Here, we calculate the decay rate for the same Lagrangian as in Eq.~(\ref{IntLang}), but this time inside a cavity for the modes of $\phi_1$ rather than in free space. The decaying field $\Phi$ is not contained inside the cavity, but if the decay process happens inside the cavity the product field $\phi_1$ particles are contained inside.  To simplify the calculation, we consider a rectangular cavity that is open along the $z$- direction and has boundaries in the $x$-  and $y$- directions. We also assume that the field $\phi_1$ is massless. These simplifications are made purely for computational ease; however, the approach can be generalized to an arbitrary cavity geometry and nonzero mass for the $\phi_1$.

The decay process can be decomposed into two parts (i) decay happening inside a cavity region and (ii) a decay happening outside the cavity region. While  the region inside the cavity the Dirichilet boundary condition ensures discrete set of transverse momentum modes and hence the density of modes in the interior region gets renegotiated, the region outside the cavity still supports a continuous distribution of modes \cite{Cernotik:2019lvz}, with the density of modes being unaltered compared to the previous case of free space. We therefore focus  on the decay rate  inside the  cavity subregion to compare the decay rate w.r.t. the corresponding free space case in the same subregion, Eq.(\ref{FiniteRegionRate}).

\subsection{Decay inside a cavity region}

 At leading order in $\lambda$ the amplitude of decay inside the cavity is still given by $\bra{\text{out}}-i\lambda\int d^4x\hat{\Phi}(x)\hat{\phi}_1(x)\hat{\phi}_1(x)\ket{\text{in}}$.
 We take  the cavity to be open along the $z$-direction. This allows us to take the field modes to be plane waves along the $z$-axis, while applying boundary conditions $\phi_1(-L,L) =0$ in the $x$- and $y$- directions yields standing wave configuration of the the Klein-Gordon equation in these directions.
 The field description for $\phi_1$ clearly shows the impact of the boundary conditions along the $x$ and $y$ directions, which results in the discretization of $k_x$ and $k_y$, 
\begin{equation}\label{equation_12}
    \hat{\phi}_1=\frac{1}{L}\sum_{\tilde{n}_1=0}^{\infty}\sum_{\tilde{m}_1=0}^{\infty}\int\frac{dk_z}{\sqrt{4 \pi\omega_{k_z,\tilde{n},\tilde{m}}}}\bigg(\hat{a}_{k_z,\tilde{n},\tilde{m}}e^{ik_z z} e^{-i \omega_{k_z,\tilde{n},\tilde{m}}t } \cos{\frac{(2\tilde{n}_1 +1)\pi x}{2L}}\cos{\frac{(2\tilde{m}_1 +1)\pi y}{2L}}+ \text{H.C.}\bigg).
\end{equation}
With the identification of odd half integer variables  ${n} =(2\tilde{n}+1)/2, {m} =(2\tilde{m}+1)/2$, the dispersion relation for the cavity modes from the Klein-Gordon equation can be written as 
\begin{equation}
\omega^2_{\{k\}} \equiv \omega_{k_z,n_1,m_1}^2={k_z}^2+\frac{\pi^2}{L^2}({n}_1^2+{m}_1^2)\,.
\end{equation}
Since the decaying field has a mode decomposition, that of the free space 
the scattering amplitude element for the decay {\it within the cavity region} can be expressed as
\begin{eqnarray}
    (-i\lambda)\int  \bra{\{k_z^{(1)},n_1,m_1\}, \{k_z^{(2)},n_2,m_2\}}\hat{\Phi}\hat{\phi}_1\hat{\phi}_1|\vec{P}_{\Phi}  \rangle d^4x
    = 
    (-i\lambda)  \nonumber\\
    \times \underbrace{\frac{1}{L}\int_{-L}^L\bigg(\cos{\frac{n_1\pi x}{L}}\cos{\frac{n_2\pi x}{L}} e^{i k_x x}\bigg) dx}_{A_{n_1,n_2,k_x}}
     \times \underbrace{\frac{1}{L}\int_{-L}^L \bigg(\cos{\frac{m_1\pi y}{L}}\cos{\frac{m_2\pi y}{L}} e^{i k_y y}\bigg) dy}_{B_{m_1,m_2,k_y}}  \nonumber\\
   \times \int_{-\infty}^\infty e^{i(k_z-k^{(2)}_z-k^{(1)}_z)z}dz\int_{-\infty}^\infty e^{i\left[\omega_{\vec{k}}-\omega_{\{k^{(1)}\}}-\omega_{\{k^{(2)}\}}\right]t}dt \nonumber\\
    = (-i\lambda)(2\pi)^2A_{n_1,n_2,k_x}B_{m_1,m_2,k_y}~\delta(k_z-k^{(2)}_z-k^{(1)}_z)\delta(\omega_{\vec{k}}-\omega_{\{k^{(1)}\}}-\omega_{\{k^{(2)}\}})\,,
\end{eqnarray}
where the in-state is depicted as $\ket{\vec{P}_\Phi} \equiv  |\vec{k}\rangle =|0,0, k_z \rangle$ and the out-states to be summed over are depicted as $ \ket{\{k_z^{(1)},n_1,m_1\}, \{k_z^{(2)},n_2,m_2\}}$  identifying the standing modes along the transverse directions and the propagating mode along the longitudinal direction of the cavity.
In the last line, the functions $A_{n_1,n_2,k_x}$ and $B_{m_1,m_2,k_y}$ arise from the integration of modes over the $x$ and $y$ directions, while the delta functions result from the integration over $z$ and $t$. If we compare this with the free space case, the result in free space contains a 4-dimensional delta function instead of a 2-dimensional one. This is expected because the presence of boundaries along $x$ and $y$ directions break translational invariance along those directions.\newline

Again, as before, we shift the computation to the co-moving Lorentz frame of the in-particle, taking $(\vec{k}=0)$  for which $A_{n_1,n_2,0} = \delta_{n_1n_2}$ and $B_{m_1,m_2,0} = \delta_{m_1m_2}$,
we obtain the probability of decay expression as
\begin{eqnarray}   \label{CavityProb}
{\cal P}_{|in\rangle \rightarrow \sum |final\rangle}   \approx \hspace{5 in}\nonumber\\
\sum_{\substack{n_1,m_1\\n_2,m_2}}\int \frac{dk_z^{(1)}}{(2\pi)2\omega_{\{k^{(1)}\}}} \int \frac{dk_z^{(2)}}{(2\pi)2\omega_{\{k^{(2)}\}}} ~\frac{\big|-i\lambda  \bra{\{k_z^{(1)},n_1,m_1\}, \{k_z^{(2)},n_2,m_2\}}\int d^4x\hat{\Phi}(x)\hat{\phi}_1(x)\hat{\phi}_1(x)\ket{\vec{P_{\Phi}}}\big|^2}{\langle \vec{P}_{\Phi}|\vec{P}_{\Phi} \rangle},
\end{eqnarray}
\begin{eqnarray}   \label{ProbT2}
   = \sum_{\substack{n_1,m_1\\n_2,m_2}}\int dk_z^{(1)} \int dk_z^{(2)} ~\frac{\lambda^2 \delta_{n_1,n_2}\delta_{m_1,m_2}~(2\pi)^2\delta(k_z-k^{(2)}_z-k^{(1)}_z)\delta(\omega_{\vec{k}}-\omega_{\{k^{(1)}\}}-\omega_{\{k^{(2)}\}})\,}{(2 \pi)^3 2\omega_{\vec{k}}\delta^3(0)~~(2\pi)2\omega_{\{k^{(1)}\}} ~(2\pi)2\omega_{\{k^{(2)}\}}} \frac{1}{L}\int_{-L}^L\bigg(\cos{\frac{n_1\pi x}{L}}\cos{\frac{n_2\pi x}{L}}\bigg) dx\nonumber\\
   \times \frac{1}{L}\int_{-L}^L \bigg(\cos{\frac{m_1\pi y}{L}}\cos{\frac{m_2\pi y}{L}} \bigg) dy  
   \times \int_{-\infty}^\infty e^{i(k_z-k^{(2)}_z-k^{(1)}_z)z}dz\int_{-\infty}^\infty e^{i\left[\omega_{\vec{k}}-\omega_{\{k^{(1)}\}}-\omega_{\{k^{(2)}\}}\right]t}dt, ~~~~~~~~~~~~~~~~~~~~~
   \end{eqnarray}
which can be written as an integral inside the cavity region
\begin{equation}   \label{ProbT2}
  {\cal P}_{|in\rangle \rightarrow \sum |final\rangle}   \approx \int dt\left[\int_{-L}^{L} dx \int_{-L}^{L} dy \int_{-\infty}^{\infty} dz\rho_{\Phi \rightarrow \phi_1 \phi_2}(x,y,z)\right]
 \end{equation}
with a decay rate density defined as
\begin{equation}
\begin{split}
    \rho_{\Phi \rightarrow \phi_1 \phi_2} (x,y,z)&= \frac{\lambda^2}{(2\pi)^3 \delta^3(0) 4ML^2}\sum_{n_1,m_1}   \cos^2\left( \frac{n_1 \pi x}{L}\right)  \cos^2\left( \frac{m_1 \pi y}{L}\right)\int\frac{dk^{(1)}_z dk^{(2)}_z}{\omega_{\{k_1\}}^2}\delta(M-2\omega_{\{k_1\}})\delta(k^{(2)}_z+ k^{(1)}_z).
    \end{split}
\end{equation}
Clearly, presence of boundaries in the transverse direction break the uniformity of the decay in those directions and leads to an integrated rate of decay inside the cavity volume as
\begin{equation}
\begin{split}
\Gamma_{V_c} &= \frac{V_c}{(2\pi)^3\delta^3(0)}\frac{\lambda^2}{4 M L^2}\sum_{n_1,m_1}\int\int\frac{dk^{(1)}_z dk^{(2)}_z}{\omega_{\{k_1\}}^2}\delta(M-2\omega_{\{k_1\}})\delta(k^{(2)}_z+ k^{(1)}_z).
    \end{split}
\end{equation}
Here as well, the prefactor $V_c/\delta^3(0)$ depicts the probability factor for a momentum eigenstate $\ket{\vec{P}_{\Phi}}$  with a uniform spatial distribution profile to undergo decay process inside the cavity region $V_c = L^2 \delta(0)$. 
Integrating out one of the decay particle's state $k_z^{(2)}$, using the dispersion relation for  the product  particle  and resorting back to $\omega_{\{k_1\}} \equiv \omega_{k_z,n_1,m_1}$, one can write $dk_z = \omega_{k_z,n_1,m_1}d\omega_{k_z,n_1,m_1}/k_z$. Transforming the integral into energy space of the decay particle, one can already see that decay  proofile develops a structure which has strong enhancement around selected points, i.e. {\it  resonance points}, 
\begin{equation} \label{CavityRate}
\begin{split}
    \Gamma_{V_c}
    &=\frac{V_c}{(2\pi)^3\delta^3(0)}\lambda^2\int_0^\infty \underbrace{\sum_{n_1,m_1} \frac{\omega_{k_z,n_1,m_1}\Theta\left(\omega_{k_z,n_1,m_1} - \frac{\pi}{L}\sqrt{(n_1^2+m_1^2)}\right) }{M L^2\sqrt{(\omega_{k_z,n_1,m_1})^2 - \frac{\pi^2}{L^2}(n_1^2+m_1^2)}}}_{\rho(\omega_{k_z,n_1,m_1})}  \underbrace{\frac{\delta\bigg(2\omega_{k_z,n_1,m_1}-M\bigg)}{4\omega_{k_z,n_1,m_1}^2}}_{{\cal K}(\omega_{k_z,n_1,m_1}; ~M)} d\omega_{k_z,n_1,m_1}.
    \end{split}
\end{equation}
The generalized form of $\rho(\omega_{k_z,n_1,m_1}){\cal K}(\omega_{k_z,n_1,m_1}; ~M)$ has a summation over the standing mode configurations which has an abrupt rise whenever the denominator vanishes. As a result, unlike  the free space the density function does not remain diluted at low frequencies any longer. We can see that density of modes has a strong sawtooth like frequency support at resonant frequencies $\omega_{k_z,n_1,m_1}\sim \frac{\pi}{L}\sqrt{(n_1^2+m_1^2)}$ \cite{Stargen:2021vtg}. This leads to a strong enhancement in the whole process whenever the mode frequency matches these resonant values for a specific value of $L$. 
Therefore, we can see that the decay  channel of $\Phi$ into $\phi_1$ gets significantly enhanced,  in a cavity environment by a precise selection of the geometry such that $\sqrt{\frac{M^2}{4}-\frac{\pi^2}{L^2}(n_1^2+m_1^2)} \longrightarrow 0$.
\begin{figure}
\begin{minipage}{.8\textwidth}
  \centering
  \includegraphics[width=.6\linewidth]{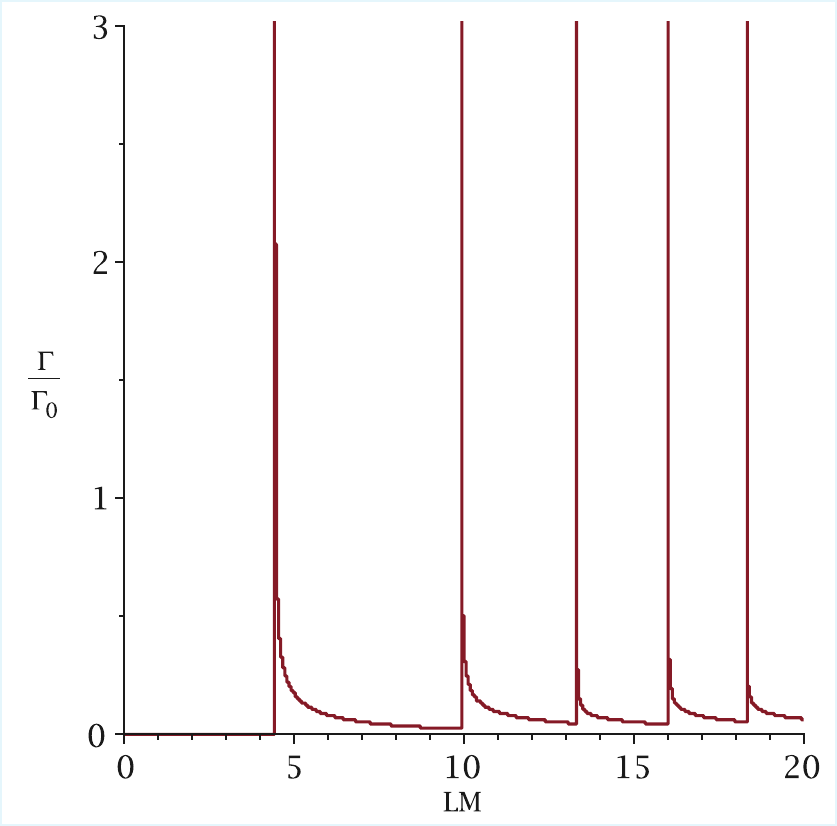}
  \caption{Plot of $\frac{\Gamma}{\Gamma_0}$ as a function of dimensionless quantity $LM$( $L$ denoting the cavity size and $M$ the mass of the decaying particle)}
  \label{fig:sfig1}
\end{minipage}%
\hspace{0.1 in}
\begin{minipage}{.5\textwidth}
  \centering
  \includegraphics[width=\linewidth]{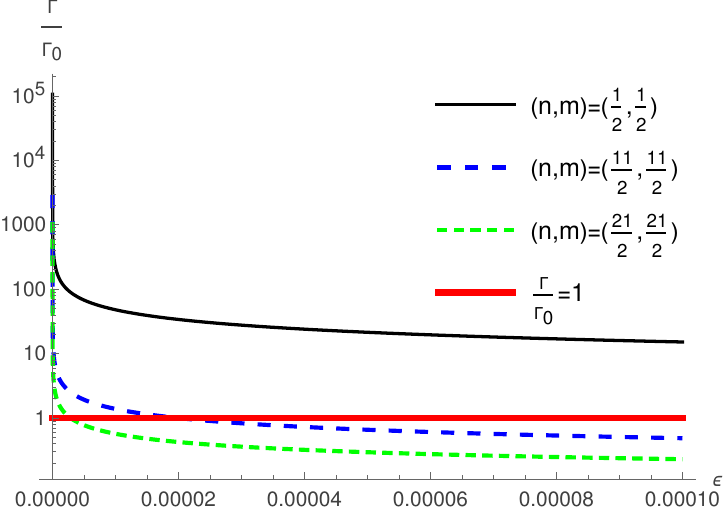}
  \caption{The width of the enhancement peak around different resonance points. Higher order resonances offer smaller width of enhancements.}
  \label{fig:sfig2}
\end{minipage}
\end{figure}
In the free space $L \rightarrow \infty$ limit, $n\pi/L = (2\tilde{n}+1)\pi/2L \rightarrow k_x^{(1)}$, $\pi \sum_n/L =\sum_n \Delta k_n \rightarrow \int dk_x$, $V_c/(2\pi)^3\delta^3(0) \rightarrow 1$ etc.,
we get (see appendix \ref{appendix_2})
\begin{equation}\label{eq_20}
\begin{split}
  \Gamma_{V_c} \rightarrow  \Gamma_0 &=\frac{\lambda^2}{4\pi^2 M^2} \int_0^\infty \int_0^\infty  {{dk_xdk_y} \over{\sqrt{\frac{M^2}{4} - k_x^2 - k_y^2}}}\Theta \bigg( \frac{M}{2}- \sqrt{ k_x^2 + k_y^2} \bigg) = \frac{\lambda^2}{16\pi M}\,,
    \end{split}
\end{equation}
We can see, the decay rate obtained in the cavity is exactly one-half of the corresponding free-space decay rate. This reduction arises from the boundary conditions imposed on the cavity, which allow only the cosine modes. In free space, both sine and cosine modes contribute to the decay process. Since the sine-mode contribution is absent in the cavity, the number of available final states is reduced by a factor of two, leading to a decay rate that is half of the free-space value. We see that the results derived in a contained environment strongly supports the decay channel if the dimensions of the cavity is suitably tailored. Comparing with the cavity-less expression of decay {\it in the same volume of space}, we can define a ratio of  enhancement  factor  as
\begin{equation}
\begin{split}
   \frac{\Gamma}{\Gamma_0} \equiv \frac{\Gamma_{V_c}}{(\Gamma_0)_{V=V_c}}   = \frac{1}{2ML^2}\sum_{n_1,m_1} \frac{ \Theta\left(M - \frac{\pi}{L}\sqrt{(n_1^2+m_1^2)}\right) }{\sqrt{M^2-\frac{\pi^2}{L^2}(n_1^2+m_1^2)}}.
    \end{split}
\end{equation}
Clearly around the points of resonance where $M L\sim \pi \sqrt{n_1^2 +m_1^2}$ the decay rate receives a significant enhancement making the ratio $\Gamma/\Gamma_0\gg 1$. Furthermore, each resonance point also offers some width for the  cavity dimension $LM$ around which the ratio remains $\Gamma/\Gamma_0\gg 1$. Thus this selection of cavity geometry offers a strong enhancement in the decay rate compared to other mechanism of enhancing the decay through external stimuli \cite{Weldon:1983jn,Coppola:2019idh}.

However, it must also be noted that the region of enhancement around  $LM \sim 2 \pi \sqrt{n_1^2 +m_1^2}$ also shrinks 
rapidly with the higher values of $LM$ (see figure  depicting the enhancement factor around various resonant points, i.e., $LM= 2 \pi \sqrt{n_1^2 +m_1^2} + \epsilon$). One can see that around the lower resonant points the  spread  of $\epsilon$ for the enhancement  region is appreciably large, which shrinks as we move on to select higher resonance point geometries. As the resonant structure is obtained  through a judicious selection of the quantity $LM$, interestingly  for low mass field decay the cavity geometry could be conveniently large enough. Therefore, the decay of lighter particles receives a large tolerance in the cavity geometry to show up the density enhanced decay rate, compared to a massive particle which needs a microscopic yet a very precise geometry to display the enhancement, challenging to achieve. 

\section{Conclusion}
In this work, we have investigated the modification of rates for physical processes inside a cavity. The  rate of any physical process is fundamentally governed by the density of states $\rho(\omega_k)$ of the product field,  which in free space scales as $\propto \omega_k^2$. Field theoretic processes therefore, in which infrared modes modes participate primarily, remain subdued due to suppression of low-frequency modes. Consequently, free-space  processes are typically  efficient  only when dominated by high-energy contributions, involving highly energetic states.  The  rate of such processes depends on mainly two ingredients, the density of modes of the product field and the contribution function. The contribution function through the transition amplitude,  depends on the interaction Lagrangian  as well as on the states participating in the process, while the density of modes solely cares about the geometric conditions employed on the product fields. In this work, we have shown that the imposition of boundary conditions through a cavity alters the density of states in a nontrivial way: the mode structure is reshaped so as to provide strong support near specific resonance frequencies, even in the infrared sector. This modification significantly enhances the influence of infrared modes, thereby dramatically enhancing the efficiency of the low energy processes, for instance in the decay dynamics. Building on this observation, we analyzed the decay of a massive particle inside a rectangular cavity and demonstrated that the decay rate exhibits a pronounced enhancement near the cavity resonance points, as shown in fig (\ref{fig:sfig1}). This geometry of the cavity, therefore, eases out the requirement of highly energetic states and provides a new avenue through which QFT effects can be studied. 

At the tree level, effect of cavity on transition amplitude appears as a consequence of breaking of Poincare symmetry. Therefore the divergences we obtain are seemingly independent of the interaction  Lagrangian and depend on cavity geometry alone and are expected to remain present even with more general forms of interaction. Though the present calculation is performed at tree level, the central result relies on the density of states $\rho(\omega_k)$, which continues to govern the behavior of the decay rate even at loop level. We therefore expect the enhancement to persist beyond tree order. Nevertheless, loop corrections may introduce additional subtleties: loop integrals must now be performed in the restricted momentum space dictated by the cavity boundary conditions, which may generate new divergence structures or modify existing ones. A detailed loop-level analysis will be presented elsewhere.


Since imposition of boundary conditions on photons through a cavity is relatively straight forward, the results obtained here have direct implications for QED processes \cite{PhysRev.76.769,PhysRev.74.1439}. In laboratory environments, a large fraction of interaction that can be directly probed are electromagnetic in nature. Any modification in photon density of states, such as the cavity induced effect as analyzed here, has the potential to substantially alter the interaction rate. Also the processes such as electron positron pair annihilation $(e^{+}+e^{-} \longleftrightarrow 2\gamma)$ or even photon induced pair creation $(\gamma+\gamma \longleftrightarrow e^{+}+e^{-})$ processes are sensitive to the available photon modes. Since these reactions rely on the emission or absorption of photons, changes in the density of states can influence both their amplitudes and their effective rates inside a cavity.
Similarly due to sensitivity with the low mass of decaying species, the results of this work can also be relevant for radiative neutrino decay
$\nu_i\rightarrow\nu_j+\gamma$ for instance, a process of continued theoretical and phenomenological interest \cite{Porto-Silva:2020gma, PhysRevD.25.766}. Neutrino radiative transitions have been studied in various environments including dense media \cite{PhysRevLett.64.1088}, astrophysical setting \cite{PhysRevD.55.7951} and more recently cosmological and laboratory context \cite{PhysRevLett.127.131102}. Our proposal offers a complementary approach in which a neutrino traverses inside a suitably engineered cavity with dimensions consistent with one of the resonance conditions $(LM)$. The modified photon density of states in this  environment can significantly enhance the radiative decay rate. This suggests that the precision-fabricated cavities could potentially serve as laboratory amplifiers for detecting extremely rare radiative processes. 

Finally, while our analysis focuses on the decay of a massive particle, the framework may be extended to massless particles. In free space, decay of a massless particle is restricted by the symmetries of the system \cite{Fiore:1995ai} which may receive significant altercation in a cavity environment. In addition to standard decay channel, our cavity enhancement mechanism may also apply to more exotic process such as decay of massive graviton into photons, $G \to \gamma +\gamma$. Massive spin-2 particle are well studied in phenomenological and extra-dimensional models and their decay has been investigated in collider experiments \cite{dEnterria:2023npy}. It  will be  worthwhile  to study the effect of reduction  of symmetries of product field  on such  processes. We plan to investigate these issues in a future work.

\section{Acknowledgments}
 Research of K.L. is partially supported by ANRF, Government of India, through a CORE research grant no. CRG/2023/004641.
The authors thank Ambresh Shivaji for useful comments.
\bibliographystyle{apsrev4-2}
\bibliography{Draft}

\appendix
\section{Decay rate calculation}\label{appendix_A}
In this appendix, we present the full tree-level decay rate calculation, detailing each step for completeness and to reproduce Eq. (\ref{equation_5}). We express the in-state as\label{equ_1}
	\begin{equation}\label{equ_1}
		\ket{in}=N\int\frac{d^3{\vec{P}}}{2E_{\vec{P}}}\tilde{\Phi}(\vec{P})\ket{\vec{P}}\,.
	\end{equation}
with $N=\frac{1}{\sqrt{\braket{in|in}}}$ is the normalization constant. The  function $\tilde{\Phi}(\vec{P})$ is the Fourier-transformed spatial wavefunction, and $\ket{\vec{P}}$ is the one-particle momentum state.
The probability that the initial particle decays into a final state consisting of two particles with momentum $\vec{p_1}$ and $\vec{p_2}$ is 
\begin{equation}\label{equ_2}
	P=\int\frac{d^3\vec{p}_1d^3\vec{p}_2}{2E_{\vec{p}_2}2E_{\vec{p}_2}}|\braket{\vec{p}_1,\vec{p}_2|\hat{T}|in}|^2\,.
\end{equation}
Here $\hat{T}=(-i\lambda)\int d^4x\hat{\Phi}(x)\hat{\phi}_1(x)\hat{\phi}_1(x)$. Now we substitute $\ket{in}$ state in equation (\ref{equ_2}) and get,
 \begin{equation}\label{equ_3}
 	P=\frac{\int\frac{d^3\vec{p}_1d^3\vec{p}_2}{2E_{\vec{p}_2}2E_{\vec{p}_2}}\int\frac{d^3{\vec{P}}}{2E_{\vec{P}}}\int\frac{d^3{\vec{P^\prime}}}{2E_{\vec{P}^{\prime}}}\tilde{\Phi}(\vec{P})\tilde{\Phi}^*(\vec{P}^\prime)\braket{\vec{p}_1,\vec{p}_2|T|\vec{P}}\braket{\vec{p}_1,\vec{p}_2|T|\vec{P}^\prime}^*}{\int\frac{d^3{\vec{P}}}{2E_{\vec{P}}}\int\frac{d^3{\vec{P^\prime}}}{2E_{\vec{P}^{\prime}}}\tilde{\Phi}(\vec{P})\tilde{\Phi}^*(\vec{P}^\prime)2E_{\vec{P}}\delta^3(\vec{P}-\vec{P^\prime})}\,.
 \end{equation}
We next compute the S-matrix element between the initial and final states.
\begin{align}
	\braket{\vec{p}_1,\vec{p}_2|\hat{T}|\vec{P}}&=\braket{\vec{p}_1,\vec{p}_2|(-i\lambda)\int d^4x\hat{\Phi}(x)\hat{\phi}_1(x)\hat{\phi}_1(x)|\vec{P}}\,\,\,\text{upto tree level}\\
	&=(-i\lambda)\delta^4(P-p_1-p_2)\,. 
\end{align}
Now putting this in equation (\ref{equ_3}) we get
\begin{align}
	P&=\lambda^2\frac{\int\frac{d^3\vec{p}_1d^3\vec{p}_2}{2E_{\vec{p}_1}2E_{\vec{p}_2}}\int\frac{d^3{\vec{P}}}{2E_{\vec{P}}}\int\frac{d^3{\vec{P^\prime}}}{2E_{\vec{P}^{\prime}}}\tilde{\Phi}(\vec{P})\tilde{\Phi}^*(\vec{P}^\prime)\delta^4(P-p_1-p_2)\delta^4(P^\prime-p_1-p_2)}{\int\frac{d^3{\vec{P}}}{2E_{\vec{P}}}\int\frac{d^3{\vec{P^\prime}}}{2E_{\vec{P}^{\prime}}}\tilde{\Phi}(\vec{P})\tilde{\Phi}^*(\vec{P}^\prime)2E_{\vec{P}}\delta^3(\vec{P}-\vec{P^\prime})}\\
	&=\lambda^2\frac{\int\frac{d^3\vec{p}_1d^3\vec{p}_2}{2E_{\vec{p}_1}2E_{\vec{p}_2}}\int\frac{d^3{\vec{P}}}{2E_{\vec{P}}}\int\frac{d^3{\vec{P^\prime}}}{2E_{\vec{P}^{\prime}}}\tilde{\Phi}(\vec{P})\tilde{\Phi}^*(\vec{P}^\prime)\delta^4(P-p_1-p_2)\delta^4(P-P^\prime)}{\int\frac{d^3{\vec{P}}}{2E_{\vec{P}}}\tilde{\Phi}(\vec{P})\tilde{\Phi}^*(\vec{P})}\\
	&=\lambda^2\frac{\int\frac{d^3\vec{p}_1d^3\vec{p}_2}{2E_{\vec{p}_1}2E_{\vec{p}_2}}\int\frac{d^3{\vec{P}}}{2E_{\vec{P}}}\int\frac{d^3{\vec{P^\prime}}}{2E_{\vec{P}^{\prime}}}\tilde{\Phi}(\vec{P})\tilde{\Phi}^*(\vec{P}^\prime)\delta^4(P-p_1-p_2)\delta^3(\vec{P}-\vec{P}^\prime)\delta(E_{\vec{P}}-E_{\vec{P}^\prime})}{\int\frac{d^3{\vec{P}}}{2E_{\vec{P}}}\tilde{\Phi}(\vec{P})\tilde{\Phi}^*(\vec{P})}\\
	&=\delta(0)\lambda^2\frac{\int\frac{d^3\vec{p}_1d^3\vec{p}_2}{2E_{\vec{p}_1}2\omega_{\vec{p}_2}}\int\frac{d^3{\vec{P}}}{2E_{\vec{P}}2E_{\vec{P}}}\tilde{\Phi}(\vec{P})\tilde{\Phi}^*(\vec{P})\delta^4(P-p_1-p_2)}{\int\frac{d^3{\vec{P}}}{2E_{\vec{P}}}\tilde{\Phi}(\vec{P})\tilde{\Phi}^*(\vec{P})}\,.
\end{align}
Since in our calculation (see section \ref{Section_II}) the decaying particle is taken to be in a definite momentum state, we consider $\tilde{\Phi}(\vec{P})=\delta^3(\vec{P}-\vec{P_\Phi})$. Upon inserting the expression for $\tilde{\Phi}$ we get the following expression

\begin{equation}
    P=\delta(0)\lambda^2\frac{\int\frac{d^3\vec{p}_1d^3\vec{p}_2}{2E_{\vec{p}_1}2E_{\vec{p}_2}}\int\frac{d^3{\vec{P}}}{2E_{\vec{P}}2E_{\vec{P}}}\delta^3(\vec{P}-\vec{P_\Phi})\delta^3(\vec{P}-\vec{P_\Phi})\delta^4(P-p_1-p_2)}{\int\frac{d^3{\vec{P}}}{2E_{\vec{P}}}\delta^3(\vec{P}-\vec{P_\Phi})\delta^3(\vec{P}-\vec{P_\Phi})}\,.
\end{equation}
Now after doing the integration over $\vec{P}$ we get 

\begin{equation}\label{equation_11}
\begin{split}
    P&=\delta(0)\lambda^2\frac{1}{2E_{\vec{P}_\Phi}}\frac{\delta^3(P_\Phi-P_\Phi)}{\delta^3(P_\Phi-P_\Phi)}\int\frac{d^3\vec{p}_1d^3\vec{p}_2}{2E_{\vec{p}_1}2E_{\vec{p}_2}}\delta^4(P_\Phi-p_1-p_2)\\
    &=\frac{\delta(0)\lambda^2}{2E_{\vec{P}_\Phi}} \frac{\delta^3(0)}{\delta^3(0)}\int\frac{d^3\vec{p}_1d^3\vec{p}_2}{2E_{\vec{p}_1}2E_{\vec{p}_2}}\delta^4(P_\Phi-p_1-p_2)\,.
    \end{split}
\end{equation}
To assign a meaningful interpretation to $\delta(0)$ we express the delta function through its integral representation as
\begin{equation}
    \delta(E_{\vec{P}}-E_{\vec{P}^\prime})=\int_{-\infty}^\infty e^{it(E_{\vec{P}}-E_{\vec{P}^\prime})}dt\,.
\end{equation} 
Now if $E_{\vec{P}}=E_{\vec{P}^\prime}$, then
\begin{equation}
    \delta(0)=\int_{-\infty}^\infty dt=\lim_{T\to\infty}\int_{-\frac{T}{2}}^{\frac{T}{2}}dt=\lim_{T\to\infty}T\,.
\end{equation} 
Similarly the three dimensional delta function can be represented as the spatial volume 
\begin{equation}
\delta^3(0) = \lim_{L\rightarrow \infty}L^3
\end{equation}    
Thus we can define a long time average decay rate per unit volume as $\rho_{\Phi \rightarrow \phi \phi}=P/\delta(0) \delta^3(0)$. Applying this to equation (\ref{equation_11}) we arrive at 
\begin{equation}
    \Gamma \equiv \int d^3 x \rho_{\Phi \rightarrow \phi \phi} = \frac{\lambda^2}{2E_{\vec{P}_\Phi}}\int\frac{d^3\vec{p}_1d^3\vec{p}_2}{2E_{\vec{p}_1}2E_{\vec{p}_2}}\delta^4(P_\Phi-p_1-p_2)\,.
\end{equation}
Since in the rest frame of the decaying particle $\Phi$ we have $E_{\vec{P}_\Phi}=M$ we can therefore write the decay rate as  in Eq. (\ref{equation_5}),
\begin{equation}
    \Gamma_0=\frac{\lambda^2}{2M}\int\frac{d^3\vec{p}_1d^3\vec{p}_2}{2E_{\vec{p}_1}2E_{\vec{p}_2}}\delta^4(P_\Phi-p_1-p_2)\,.
\end{equation}

\section{Free-space limit of the quantities defined within the cavity}\label{appendix_2}
In this appendix we obtain the free space limits of cavity results and recover the standard free space expressions for the fields and decay rate. In the free–space limit \(L \rightarrow \infty\), the discrete momenta 
\begin{equation}
\begin{split}
\frac{n\pi}{L} = \frac{(2\tilde{n}+1)\pi}{2L} \;\longrightarrow\; k_x^{(1)},
\qquad 
\frac{\pi}{L}\sum_{n} = \sum_{n} \Delta k_n \;\longrightarrow\; \int dk_x,
\end{split}
\end{equation}
With these free-space limits, 
 the decay rate expression can be evaluated by carrying out the integration over $k_x$ and $k_y$ in equation (\ref{eq_20}).
\begin{align}
    \Gamma=\frac{\lambda^2}{4\pi^2M^2}\int_0^\infty\int_0^\infty \frac{dk_x dk_y\Theta \bigg( \frac{M}{2}- \sqrt{ k_x^2 + k_y^2} \bigg)}{\sqrt{\frac{M^2}{4}-k_x^2-k_y^2}}\,.
\end{align}
The Heaviside function restricts the integration to the quarter disk
\(
k_x^2 + k_y^2 \le (M/2)^2
\)
in the first quadrant.  
Introducing polar coordinates in momentum space:
\begin{align}
k_x = k\cos\theta_k,
\qquad
k_y = k\sin\theta_k,
\qquad
k\ge0,\quad 0\le\theta_k\le\frac{\pi}{2},
\end{align}
with Jacobian \(dk_x\,dk_y = k\,dk\,d\theta_k\).
Then
\begin{align}
\sqrt{\frac{M^2}{4}-k_x^2-k_y^2}
=\sqrt{\frac{M^2}{4}-k^2},
\qquad
0\le k\le \frac{M}{2}.
\end{align}
Thus the integral becomes
\begin{align}
	\Gamma
	&= \,\frac{\lambda^2}{2\pi^2 M^2}
	\int_{0}^{\pi/2}\!\! d\theta_k
	\int_{0}^{M/2}
	\frac{k\,dk}{\sqrt{\frac{M^2}{4}-k^2}}.
\end{align}
The angular integral gives a factor of \(\pi/2\):
\begin{align}
\Gamma
= \,\frac{\lambda^2}{4\pi^2 M^2}\cdot\frac{\pi}{2}
\int_{0}^{M/2}\frac{k\,dk}{\sqrt{\frac{M^2}{4}-k^2}}=\frac{\lambda^2}{16\pi M}.
\end{align}

As shown in our calculation the $\rho(\omega_{k_z,n,m})$ plays a central role. In this appendix, we also derive its free space limit starting from the cavity expression, so that we recover the  free-space form $\rho({\omega_{\vec{k}}})$. We begin with equation (\ref{CavityRate}).
In free space limit $L\to\infty$ equation (\ref{CavityRate}) becomes
\begin{equation}
    \Gamma=\lambda^2\int_0^\infty d\omega\int_0^\infty\int_0^\infty dk_x dk_y\frac{\omega\Theta\big(\omega-\sqrt{k_x^2+k_y^2}\big)}{\pi^2 M\sqrt{\omega^2-k_x^2-k_x^2}}\frac{\delta(M-2\omega)}{4\omega^2}\,.
\end{equation}
We can perform the integration over $k_x$ and $k_y$ which yields. 
\begin{equation}\label{B_13}
    \Gamma=\lambda^2\int_0^\infty d\omega\underbrace{\frac{2\pi\omega^2}{(2\pi)^2 M}}_{\rho(\omega)}\underbrace{\frac{\delta(M-2\omega)}{4\omega^2}}_{\mathcal{K}(\omega,M)}\,.
\end{equation} 
We can see that equation (\ref{B_13}) agrees with the equation (\ref{eq_8}) in section \ref{Section_II} when the initial particle decays into massless final states.

\section {Dimensional Analysis and Lorentz Invariance}
$$ S = \lambda \int d^4 x \Phi \phi_1\phi_1$$
is dimensionless (in natural units)
Thus $[\Phi] =[\phi] = [\lambda] = L^{-1}$
\begin{equation}
  \lambda \bra{\vec{p_1} \vec{p_2}}\int d^4x\hat{\Phi}(x)\hat{\phi}_1(x)\hat{\phi}_1(x)\ket{\vec{P}_{\Phi}} =\lambda\delta^4(P_{\Phi}-p_1-p_2)
   \end{equation}
   
 will have the dimension of that of  ${\langle \vec{P}_{\Phi}|\vec{P}_{\Phi} \rangle^{1/2}}$ and 
 ${\langle \vec{p}_{1}\vec{p}_{2}|\vec{p}_{1}\vec{p}_{2} \rangle^{1/2}}$  multiplied.  Since 
 $\ket{\vec{P}} = \sqrt{(2\pi)^32 \omega_{\vec{P}}}\hat{a}^{\dagger}_{\vec{P}}\ket{0}_{\Phi}$
and $|\vec{p}_1,\vec{p}_2\rangle  =   \sqrt{(2\pi)^32\omega_{\vec{p}_1}}\sqrt{(2\pi)^32\omega_{\vec{p}_2}}\hat{a}_{\vec{p}_1}^{\dagger} \hat{a}_{\vec{p}_2}^{\dagger}\ket{0}_{\phi_1}\ket{0}_{\phi_1}$. The dimensionality of the state $\ket{\vec{P}}$ can be obtained from $[\langle \vec{P}\ket{\vec{P}}] = [2 \omega_{\vec{P}} ~~{}_{\Phi}\bra{0} \hat{a}_{\vec{P}}\hat{a}^{\dagger}_{\vec{P}} \ket{0}_{\Phi}] =  [2 \omega_{\vec{P}}~~\delta^3(0)] \sim L^2.$ Similarly $\langle \vec{p}_{1}\vec{p}_{2}|\vec{p}_{1}\vec{p}_{2} \rangle] \sim L^4$. Therefore,
\begin{equation}   
  {\cal P}_{|in\rangle \rightarrow \sum |final\rangle}   \approx \underbrace{\int \frac{d^3\vec{p_1}}{2E_{p_1}}}_{L^{-2}}\underbrace{\int \frac{d^3\vec{p_2}}{2E_{p_2}}}_{L^{-2}} \underbrace{ \frac{~|-i\lambda\bra{\vec{p_1} \vec{p_2}}\int d^4x\hat{\Phi}(x)\hat{\phi}_1(x)\hat{\phi}_1(x)\ket{\vec{P_{\Phi}}}|^2}{\langle \vec{P}_{\Phi}|\vec{P}_{\Phi} \rangle }}_{dim \langle \vec{p}_{1}\vec{p}_{2}|\vec{p}_{1}\vec{p}_{2} \rangle = L^4}
 \end{equation}
 is dimensionless.\\
For the cavity case the   only  change in  the  analysis  is the change in the out-state $\ket{\{k_z^{(1)},n_1,m_1\}, \{k_z^{(2)},n_2,m_2\}} =  \sqrt{(2\pi)2\omega_{k_z^{(1)},n_1,m_1}}\sqrt{(2\pi)2\omega_{k_z^{(2)},n_2,m_2}}\hat{a}_{k_z^{(1)},n_1,m_1}^{\dagger}\hat{a}_{k_z^{(2)},n_2,m_2}^{\dagger}\ket{0}_{\phi_1}\ket{0}_{\phi_1}$  with a  dimensionality 
\begin{eqnarray}
[\langle \{k_z^{(1)},n_1,m_1\}, \{k_z^{(2)},n_2,m_2\} \ket{\{k_z^{(1)},n_1,m_1\}, \{k_z^{(2)},n_2,m_2\}}] = [4 \omega_{k_z^{(1)},n_1,m_1}\omega_{k_z^{(2)},n_2,m_2}  \nonumber\\ 
{}_{\phi_1}\bra{0}\hat{a}_{k_z^{(1)},n_1,m_1}\hat{a}^{\dagger}_{k_z^{(1)},n_1,m_1}  \hat{a}_{k_z^{(2)},n_2,m_2}\hat{a}^{\dagger}_{k_z^{(2)},n_2,m_2} \ket{0}_{\phi_1}] \sim L^0.
\end{eqnarray}
Thus the state  is  dimensionless. The dimension of the amplitude
\begin{eqnarray}
    \underbrace{(\lambda)}_{L^{-1}}\int  \underbrace{\bra{\{k_z^{(1)},n_1,m_1\}, \{k_z^{(2)},n_2,m_2\}}}_{L^0}\underbrace{\hat{\Phi}\hat{\phi}_1\hat{\phi}_1}_{L^{-3}}\underbrace{|\vec{P}_{\Phi}  \rangle}_{L^{1}} \underbrace{d^4x}_{L^4} \nonumber\\
    \sim  \underbrace{(\lambda)}_{L^{-1}} \underbrace{A_{n_1,n_2,k_x}B_{m_1,m_2,k_y}}_{L^0}~\underbrace{\delta(k_z-k^{(2)}_z-k^{(1)}_z)\delta(\omega_{\vec{k}}-\omega_{\{k^{(1)}\}}-\omega_{\{k^{(2)}\}})}_{L^2}  \sim L\,,
\end{eqnarray}
making Eq.(\ref{CavityProb}) dimensionless.
\end{document}